\documentclass{article}

\usepackage{hyperref}
\usepackage{amsfonts,amsmath,amssymb,amscd,amsthm}
\usepackage{latexsym}
\usepackage[all]{xy}
\usepackage{graphicx}
\usepackage{epsfig}
\usepackage{tikz}
\usetikzlibrary{matrix,arrows}

\newcommand{\R}{{\mathbb R}}

\newcommand{\N}{{\mathbb N}}

\newcommand{\dom}{{\rm dom}}

\renewcommand{\to}{\rightarrow}

\newcommand{\maps}{\colon}

\newtheorem*{theorem*}{Theorem}
\newtheorem*{definition*}{Definition}
\newtheorem*{lemma*}{Lemma}
\newtheorem*{corollary*}{Corollary}
\newtheorem*{proposition*}{Proposition}
\newtheorem*{example*}{Example}
\newtheorem*{conjecture*}{Conjecture}
\newtheorem*{remark*}{Remark}
\newtheorem*{notation*}{Notation}
\newtheorem*{convention*}{Convention}

\newdir{ >}{{}*!/-7pt/@{>}}
\newdir{> >}{*!/0pt/@{>}*!/-5pt/@{>}}

\begin{document}
\sloppy
     \title{Algorithmic Thermodynamics} 
     \author{{\small John C.\ Baez} \\
     {\small Department of Mathematics,  University of California} \\
     {\small Riverside, California 92521, USA} \\
     \\  
    {\small Mike Stay} \\
    {\small Computer Science Department, University of Auckland} \\
    {\small \it and} \\
    {\small Google, 1600 Amphitheatre Pkwy} \\
    {\small Mountain View, California 94043, USA}
    \\
    \\ {\small email: baez@math.ucr.edu, stay@google.com}
    \\
    }

\date{\today}
\maketitle

\begin{abstract}
\noindent
Algorithmic entropy can be seen as a special case of entropy as studied
in statistical mechanics.  This viewpoint allows us to apply many
techniques developed for use in thermodynamics to the subject of
algorithmic information theory.  In particular, suppose we fix a
universal prefix-free Turing machine and let $X$ be the set of
programs that halt for this machine.  Then we can regard $X$ as a set
of `microstates', and treat any function on $X$ as an `observable'.  For
any collection of observables, we can study the Gibbs ensemble that
maximizes entropy subject to constraints on expected values of these
observables.  We illustrate this by taking the log runtime, length,
and output of a program as observables analogous to the energy $E$,
volume $V$ and number of molecules $N$ in a container of gas.  The
conjugate variables of these observables allow us to define quantities
which we call the `algorithmic temperature' $T$, `algorithmic
pressure' $P$ and `algorithmic potential' $\mu$, since they are
analogous to the temperature, pressure and chemical potential.  We
derive an analogue of the fundamental thermodynamic relation $d E = T
dS - P d V + \mu d N$, and use it to study thermodynamic cycles
analogous to those for heat engines.  We also investigate the values
of $T, P$ and $\mu$ for which the partition function converges.  
At some points on the boundary of this domain of convergence, the 
partition function becomes uncomputable.  Indeed, at these points 
the partition function itself has nontrivial algorithmic entropy.
\end{abstract}

\section{Introduction}\label{intro}

Many authors \cite{BGLVZ, Chaitin1975, FT1982, Kolmogorov1965, 
LevinZvonkin, Solomonoff1964, Szilard1929, Tadaki2008} have discussed 
the analogy between algorithmic entropy and entropy as defined in 
statistical mechanics: that is, the entropy of a probability measure 
$p$ on a set $X$.  It is perhaps insufficiently appreciated that 
algorithmic entropy can be seen as a \textit{special case} of the 
entropy as defined in statistical mechanics.  We describe how to do 
this in Section \ref{entropy}.

This allows all the basic techniques of thermodynamics to be imported
to algorithmic information theory.  The key idea is to take $X$ to be
some version of `the set of all programs that eventually halt and
output a natural number', and let $p$ be a Gibbs ensemble on $X$.  A
Gibbs ensemble is a probability measure that maximizes entropy subject
to constraints on the mean values of some observables---that is,
real-valued functions on $X$.

In most traditional work on algorithmic entropy, the relevant
observable is the length of the program.  However, much of the
interesting structure of thermodynamics only becomes visible when we
consider several observables.  When $X$ is the set of programs that
halt and output a natural number, some other important observables
include the output of the program and logarithm of its runtime.  So,
in Section \ref{thermodynamics} we illustrate how ideas from
thermodynamics can be applied to algorithmic information theory using
these three observables.  

To do this, we consider a Gibbs ensemble of programs which maximizes
entropy subject to constraints on:
\begin{itemize}
\item
$E$, the expected value of the logarithm of the program's
runtime (which we treat as analogous to the energy
of a container of gas),
\item 
$V$, the expected value of the length of the program
(analogous to the volume of the container),
and 
\item
$N$, the expected value of the program's output
(analogous to the number of molecules in the gas).
\end{itemize}
This measure is of the form
\[       p = \frac{1}{Z} e^{-\beta E(x) -\gamma V(x) - \delta N(x)} \]
for certain numbers $\beta, \gamma, \delta$, where the normalizing factor
\[     Z = \sum_{x \in X} e^{-\beta E(x) -\gamma V(x) - \delta N(x)} \]
is called the `partition function' of the ensemble.  The partition
function reduces to Chaitin's number $\Omega$ when $\beta = 0$,
$\gamma = \ln 2$ and $\delta = 0$.  This number is uncomputable
\cite{Chaitin1975}.  However, we show that the partition function $Z$
is computable when $\beta > 0$, $\gamma \ge \ln 2$, and $\delta \ge 0$.

We derive an algorithmic analogue of the basic thermodynamic relation
\[          dE = T dS - P dV + \mu dN . \]
Here:
\begin{itemize}
\item 
$S$ is the entropy of the Gibbs emsemble,
\item
$T = 1/\beta$ is the `algorithmic temperature' (analogous to the
temperature of a container of gas).  Roughly speaking, this counts how
many times you must double the runtime in order to double the number
of programs in the ensemble while holding their mean length and output
fixed.
\item
$P = \gamma/\beta$ is the `algorithmic pressure' (analogous to
pressure).  This measures the tradeoff between runtime and length.
Roughly speaking, it counts how much you need to decrease the mean
length to increase the mean log runtime by a specified amount, while
holding the number of programs in the ensemble and their mean output
fixed.
\item
$\mu = -\delta/\beta$ is the `algorithmic potential' (analogous to
chemical potential).  Roughly speaking, this counts how much the mean
log runtime increases when you increase the mean output while holding
the number of programs in the ensemble and their mean length fixed.
\end{itemize}

Starting from this relation, we derive analogues of Maxwell's
relations and consider thermodynamic cycles such as the Carnot cycle
or Stoddard cycle.  For this we must introduce concepts of `algorithmic 
heat' and `algorithmic work'.

\begin{center}
\includegraphics[scale=0.15, angle=0.3]{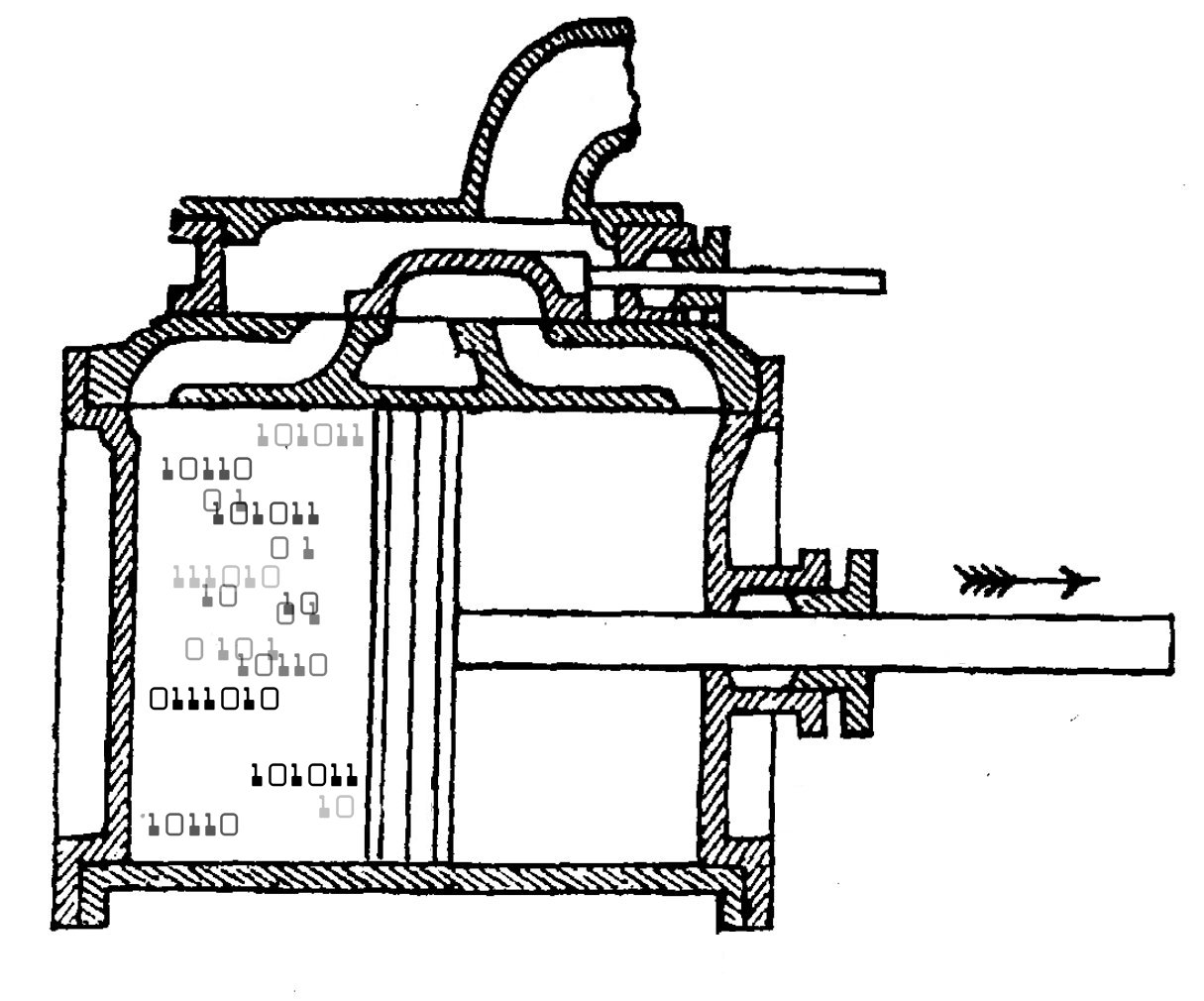}
\end{center}

Charles Babbage described a computer powered by a steam engine;
we describe a heat engine powered by programs!  We admit
that the significance of this line of thinking remains a bit 
mysterious.  However, we hope it points the way toward a further 
synthesis of algorithmic information theory and thermodynamics.
We call this hoped-for synthesis `algorithmic thermodynamics'.

\section{Related Work}

Li and Vit\'anyi use the term `algorithmic thermodynamics' for describing
physical states using a universal prefix-free Turing machine $U$.  
They look at the smallest program $p$ that outputs a 
description $x$ of a particular microstate to some 
accuracy, and define the physical entropy to be
\[ S_A(x) = (k \ln 2)(K (x) + H_x), \]
where $K(x) = |p|$ and $H_x$ embodies the uncertainty in the actual state given $x$.
They summarize their own work and subsequent work by others in chapter 
eight of their book \cite{LV}.  Whereas they consider $x=U(p)$ to be a
microstate, we consider $p$ to be the microstate and $x$ the value of
the observable $U$.  Then their observables $O(x)$ become observables
of the form $O(U(p))$ in our model.
 
Tadaki \cite{Tadaki2002} generalized Chaitin's number $\Omega$ to a function 
$\Omega^D$ and showed that the value of this function 
is compressible by a factor of exactly $D$ when $D$ is computable.
Calude and Stay \cite{CSNatural2006} pointed out that this generalization 
was formally equivalent to the partition function of a statistical
mechanical system where temperature played the role of the compressibility
factor, and studied various observables of such a system. 
Tadaki \cite{Tadaki2008} then explicitly constructed a system with
that partition function: given a total length $E$ and number of programs
$N,$ the entropy of the system is the log of the number of $E$-bit strings 
in $\dom(U)^N.$  The temperature is
\[ \frac{1}{T} = \left.\frac{\Delta E}{\Delta S}\right|_N. \]
In a follow-up paper \cite{Tadaki2009}, Tadaki showed that various other 
quantities like the free energy shared the same compressibility properties
as $\Omega^D$.  In this paper, we consider multiple variables, which is
necessary for thermodynamic cycles, chemical reactions, and so forth.

Manin and Marcolli \cite{MM2009} derived similar results in a broader context
and studied phase transitions in those systems.  
Manin \cite{ManinRenorm1, ManinRenorm2} 
also outlined an ambitious program to
treat the infinite runtimes one finds in undecidable problems as singularities
to be removed through the process of renormalization.  In a
manner reminiscent of hunting for the proper definition of the 
``one-element field'' $F_{un},$ he collected ideas from
many different places and considered how they all touch on this central theme.
While he mentioned a runtime cutoff as being analogous to an energy cutoff,
the renormalizations he presented are uncomputable.  In this paper, we
take the log of the runtime as being analogous to the energy; the randomness
described by Chaitin and Tadaki then arises as the infinite-temperature
limit.

\section{Algorithmic Entropy}\label{entropy}

To see algorithmic entropy as a special case of the entropy of a
probability measure, it is useful to follow Solomonoff
\cite{Solomonoff1964} and take a Bayesian viewpoint.  In 
Bayesian probability theory, we always start with a probability
measure called a `prior', which describes our assumptions 
about the situation at hand before we make any further observations.
As we learn more, we may update this prior.  This approach suggests
that we should define the entropy of a probability measure
\textit{relative to another probability measure}---the prior.

A probability measure $p$ on a finite set $X$ is simply a function $p
\maps X \to [0,1]$ whose values sum to 1, and its entropy is defined
as follows:
\[  S(p) = -\sum_{x \in X} p(x) \ln p(x)  .\]
But we can also define the entropy of $p$ relative to another 
probability measure $q$:
\[  S(p,q) = -\sum_{x \in X} p(x) \ln\frac{p(x)}{q(x)} .\]
This {\bf relative entropy} has been extensively studied and goes by
various other names, including `Kullback--Leibler divergence' \cite{KL}
and `information gain' \cite{Renyi}.  

The term `information gain' is nicely descriptive.  Suppose we 
initially assume the outcome of an experiment is distributed according to 
the probability measure $q$.  Suppose we then repeatedly do the experiment
and discover its outcome is distributed according to the measure $p$.
Then the information gained is $S(p,q)$.

Why?  We can see this in terms of coding.  Suppose $X$ is a
finite set of signals which are randomly emitted by some source.
Suppose we wish to encode these signals as efficiently as possible in
the form of bit strings.  Suppose the source emits the signal $x$ with
probability $p(x)$, but we erroneously believe it is emitted with
probability $q(x)$.  Then $S(p,q)/\ln 2$ is the expected extra
message-length per signal that is required if we use a code that is
optimal for the measure $q$ instead of a code that is optimal for the
true measure, $p$.

The ordinary entropy $S(p)$ is, up to a constant, just the
relative entropy in the special case where the prior assigns 
an equal probability to each outcome.  In other words:
\[  S(p) = S(p,q_0) + S(q_0)  \]
when $q_0$ is the so-called `uninformative prior', with $q_0(x) =
1/|X|$ for all $x \in X$.  

We can also define relative entropy when the set $X$ is countably
infinite.  As before, a probability measure on $X$ is a function $p
\maps X \to [0,1]$ whose values sum to 1.  And as before, if $p$ and
$q$ are two probability measures on $X$, the entropy of $p$ relative
to $q$ is defined by
\begin{equation}
\label{relative_entropy}
  S(p,q) = -\sum_{x \in X} p(x) \, \ln\frac{p(x)}{q(x)} .
\end{equation} 
But now the role of the prior becomes more clear, because there
is no probability measure that assigns the same value to each outcome!

In what follows we will take $X$ to be---roughly speaking---the
set of all programs that eventually halt and output a natural 
number.  As we shall see, while this set is countably infinite, there 
are still some natural probability measures on it, which we may take
as priors.

To make this precise, we recall the concept of a universal
prefix-free Turing machine.  In what follows we use {\bf string} to
mean a bit string, that is, a finite, possibly empty, list of 0's and
1's.  If $x$ and $y$ are strings, let $x||y$ be the concatenation of
$x$ and $y.$ A \textbf{prefix} of a string $z$ is a substring
beginning with the first letter, that is, a string $x$ such that $z =
x||y$ for some $y$.  A \textbf{prefix-free} set of strings is one in
which no element is a prefix of any other.  The \textbf{domain}
$\dom(M)$ of a Turing machine $M$ is the set of strings that cause $M$
to eventually halt.  We call the strings in $\dom(M)$
\textbf{programs}.  We assume that when the $M$ halts on the program
$x$, it outputs a natural number $M(x)$.  Thus we may think of the
machine $M$ as giving a function $M \maps \dom(M) \to \N$.

A \textbf{prefix-free Turing machine} is one whose halting programs
form a prefix-free set.  A prefix-free machine $U$ is {\bf universal}
if for any prefix-free Turing machine $M$ there exists a constant
$c$ such that for each string $x$, there exists a string $y$ with
\[ U(y) = M(x) \; \mbox{ and } \; |y| < |x| + c. \]

Let $U$ be a universal prefix-free Turing machine.  Then 
we can define some probability measures on $X = \dom(U)$
as follows.   Let 
\[  |\cdot | \maps X \to \N \]
be the function assigning to each bit string its length.  
Then there is for any constant $\gamma > \ln 2$ a probability measure 
$p$ given by
\[   p(x) = \frac{1}{Z} e^{-\gamma |x|}. \]
Here the normalization constant $Z$ is chosen to make the numbers
$p(x)$ sum to 1:
\[   Z = \sum_{x \in X} e^{-\gamma |x|}  .\]
It is worth noting that for computable real numbers $\gamma \ge \ln 2$, 
the normalization constant $Z$ is uncomputable \cite{Tadaki2002}.
Indeed, when $\gamma = \ln 2$, $Z$ is Chaitin's famous number $\Omega$.
We return to this issue in Section \ref{computability}.

Let us assume that each program prints out some natural
number as its output.   Thus we have a function
\[ N \maps X \to \N \]
where $N(x)$ equals $i$ when program $x$ prints out the number $i$.
We may use this function to `push forward' $p$ to a probability measure
$q$ on the set $\N$.  Explicitly:
\[ 
   q(i) = \displaystyle {\sum_{x \in X : N(x) = i}} e^{-\gamma |x|} \; .
\]
In other words, if $i$ is some natural number, $q(i)$ is the 
probability that a program randomly chosen according to the measure
$p$ will print out this number.  

Given any natural number $n$, there is a probability measure
$\delta_n$ on $\N$ that assigns probability 1 to this number:
\[   \delta_n(m) = \left\{ \begin{array}{cl} 1 & \textrm{if } m = n \\
                              0 & \textrm{otherwise.}  
           \end{array} \right.
\]
We can compute the entropy of $\delta_n$ relative to $q$:
\begin{equation}
\label{relative.entropy}
\begin{array}{ccl} S(\delta_n,q) &=& 
\displaystyle{ -\sum_{i \in \N}  \delta_n(i) \, 
\ln \frac{\delta_n(i)}{q(i)}}   \\
\\ &=& \displaystyle{ -\ln \left( \sum_{x \in X \colon N(x) = n} 
e^{-\gamma |x|} \right) + \ln Z .}  \\
\end{array}
\end{equation}
Since the quantity $\ln Z$ is independent of the number $n$, and
uncomputable, it makes sense to focus attention on the other part
of the relative entropy:
\[  \displaystyle{ -\ln \left( \sum_{x \in X \colon N(x) = n} 
e^{-\gamma |x|} \right) .}  
\]
If we take $\gamma = \ln 2$, this is precisely the \textbf{algorithmic
entropy} \cite{Chaitin1976,LevinZvonkin} of the number $n$.  So, up to the
additive constant $\ln Z$, we have seen that \textit{algorithmic entropy 
is a special case of relative entropy}.  

One way to think about entropy is as a measure of surprise: if you 
can predict what comes next---that is, if you have a program that 
can compute it for you---then you are not surprised.  For example, 
the first 2000 bits of the
binary fraction for 1/3 can be produced with this short Python
program:
\begin{center}
  {\tt print "01" * 1000}
\end{center}
But if the number is complicated, if every bit is surprising and
unpredictable, then the shortest program to print the number does not
do any computation at all! It just looks something like
\begin{center}
  {\tt print "101000011001010010100101000101111101101101001010"}
\end{center}
Levin's coding theorem \cite{Levin1974} says that the difference between
the algorithmic entropy of a number and its \textbf{Kolmogorov 
complexity}---the length of the shortest program that outputs it---is 
bounded by a constant that only depends on the programming
language.

So, we are seeing here that up to some error bounded by a constant,
\textit{Kolmogorov complexity is information gain}: the
information gained upon learning a number, if our prior assumption was
that this number is the output of a program randomly
chosen according to the measure $p$ where $\gamma = \ln 2$.

More importantly, we have seen that algorithmic entropy is not just 
\textit{analogous} to entropy as defined in statistical mechanics: it 
is a \textit{special case}, as long as we take seriously the Bayesian 
philosophy that entropy should be understood as relative entropy.  
This realization opens up the possibility of taking many familiar concepts 
from thermodynamics, expressed in the language of statistical mechanics,
and finding their counterparts in the realm of algorithmic information
theory.

But to proceed, we must also understand more precisely the role of the
measure $p$.  In the next section, we shall see that this type of
measure is already familiar in statistical mechanics: it is a Gibbs
ensemble.

\section{Algorithmic Thermodynamics}\label{thermodynamics}

Suppose we have a countable set $X$, finite or
infinite, and suppose $C_1, \dots, C_n \maps X \to \R$ is some
collection of functions.  Then we may seek a probability measure $p$
that maximizes entropy subject to the constraints that the mean value
of each observable $C_i$ is a given real number $\overline{C}_i$:
\[            \sum_{x \in X} p(x) \, C_i(x) = \overline{C}_i  .\]
As nicely discussed by Jaynes \cite{Jaynes1957,Jaynes2003}, the solution, 
if it exists, is the so-called {\bf Gibbs ensemble}:
\[             p(x) = \frac{1}{Z} e^{-(s_1 C_1(x) + \cdots + s_n C_n(x))}  \]
for some numbers $s_i \in \R$ depending on the desired mean values 
$\overline{C}_i$.  Here the normalizing factor $Z$ is called the 
{\bf partition function}:
\[             Z = \sum_{x \in X} e^{-(s_1 C_1(x) + \cdots + s_n C_n(x))} 
\;.  \]

In thermodynamics, $X$ represents the set of {\bf microstates} of some
physical system.  A probability measure on $X$ is also known as an
{\bf ensemble}.  Each function $C_i \maps X \to \R$ is called an {\bf
observable}, and the corresponding quantity $s_i$ is called the {\bf
conjugate variable} of that observable.  For example, the conjugate of
the energy $E$ is the inverse of temperature $T$, in units where
Boltzmann's constant equals 1.  The conjugate of the volume $V$---of
a piston full of gas, for example---is the pressure $P$ divided by
the temperature.  And in a gas containing molecules of various types,
the conjugate of the number $N_i$ of molecules of the $i$th type is
minus the `chemical potential' $\mu_i$, again divided by temperature.
For easy reference, we list these observables and their conjugate
variables below.

\begin{center}
\renewcommand{\arraystretch}{2.3}
\begin{tabular}{c|c}	
\multicolumn{2}{c}{\bf{THERMODYNAMICS}} \\
\hline
Observable    &  Conjugate Variable\\ 
\hline
energy: $E$   &  $\displaystyle{\frac{1}{T}}$ \\
volume: $V$   &  $\displaystyle{\frac{P}{T}}$  \\
number: $N_i$ &  $\displaystyle{-\frac{\mu_i}{T}}$   \\
\end{tabular}
\end{center}
\renewcommand{\arraystretch}{1}

Now let us return to the case where $X=\dom(U)$.  Recalling that programs
are bit strings, one important observable for programs is the length:
\[  |\cdot | \maps X \to \N .\]
We have already seen the measure
\[   p(x) = \frac{1}{Z} e^{-\gamma |x|} .\]
Now its significance should be clear!   This is the 
probability measure on programs
that maximizes entropy subject to the constraint that the mean length
is some constant $\ell$:
\[    \sum_{x \in X}  p(x) \, |x| = \ell .  \]
So, $\gamma$ is the conjugate variable to program length.

There are, however, other important observables that can be defined
for programs, and each of these has a conjugate quantity.  To make the
analogy to thermodynamics as vivid as possible, let us arbitrarily
choose two more observables and treat them as analogues of energy and
the number of some type of molecule.  Two of the most obvious
observables are `output' and `runtime'.  Since Levin's computable
complexity measure \cite{Levin1973} uses the logarithm of runtime as a
kind of `cutoff' reminiscent of an energy cutoff in renormalization,
we shall arbitrarily choose the log of the runtime to be analogous to
the energy, and denote it as
\[    E \maps X \to [0,\infty) \]
Following the chart above, we use $1/T$ to stand for the variable
conjugate to $E$.   We arbitrarily treat the output of a
program as analogous to the number of a certain kind of molecule, and
denote it as
\[    N \maps X \to \N . \]
We use $-\mu/T$ to stand for the conjugate variable of $N$.
Finally, as already hinted, we denote program length as
\[    V \maps X \to \N   \]
so that in terms of our earlier notation, $V(x) = |x|$.  We use
$P/T$ to stand for the variable conjugate to $V$.

\begin{center}
\renewcommand{\arraystretch}{2.3}
\begin{tabular}{c|c}	
\multicolumn{2}{c}{\bf{ALGORITHMS}} \\
\hline
Observable  & Conjugate Variable\\ 
\hline
log runtime: $E$  & $\displaystyle{\frac{1}{T}}$ \\
length:      $V$  & $\displaystyle{\frac{P}{T}}$  \\
output:      $N$  & $\displaystyle{-\frac{\mu}{T}}$   \\
\end{tabular}
\end{center}
\renewcommand{\arraystretch}{1}

Before proceeding, we wish to emphasize that the analogies here 
were chosen somewhat arbitrarily.  They are merely meant to illustrate
the application of thermodynamics to the study of algorithms.  There
may or may not be a specific `best' mapping between observables
for programs and observables for a container of gas!  Indeed, 
Tadaki \cite{Tadaki2008} has explored another analogy, where
length rather than log run time is treated as the analogue of 
energy.  There is nothing wrong with this.  However, he did not 
introduce enough other observables to see the whole structure of
thermodynamics, as developed in Sections \ref{elementary}-\ref{cycles}
below.  

Having made our choice of observables, we define the partition function
by 
\[     Z = \sum_{x \in X} e^{-\frac{1}{T}(E(x) + P V(x) - \mu N(x))} \; .\]
When this sum converges, we can define a probability measure on $X$,
the Gibbs ensemble, by
\[    p(x) = \frac{1}{Z} e^{-\frac{1}{T}(E(x) + P V(x) - \mu N(x))} \; .\]
Both the partition function and the probability measure are
functions of $T, P$ and $\mu$.  From these we can compute the mean
values of the observables to which these variables are conjugate:
\[
\begin{array}{ccc} 
    \overline{E} &=& \displaystyle{\sum_{x \in X}} p(x) \, E(x)   \\
                                                   \\
    \overline{V} &=& \displaystyle{\sum_{x \in X}} p(x) \, V(x)   \\
                                                   \\
    \overline{N} &=& \displaystyle{\sum_{x \in X}} p(x) \, N(x)   
\end{array}
\]
In certain ranges, the map $(T,P,\mu) \mapsto (\overline{E},
\overline{V}, \overline{N})$ will be invertible.  This allows us to
alternatively think of $Z$ and $p$ as 
functions of $\overline{E}, \overline{V},$ and $\overline{N}$.
In this situation it is typical to abuse language by omitting the
overlines which denote `mean value'.

\subsection{Elementary Relations} \label{elementary}

The entropy $S$ of the Gibbs ensemble is given by
\[      S = - \sum_{x \in X} p(x)\, \ln p(x) .\]
We may think of this as a function of $T, P$ and $\mu$, or 
alternatively---as explained above---as functions of 
the mean values $E, V,$ and $N$.  Then simple calculations,
familiar from statistical mechanics \cite{Reif}, show that

\begin{equation}
\label{derivative1}
\displaystyle{\left.\frac{\partial S}{\partial E}\right|_{V,N}}  =
\displaystyle{\frac{1}{T}} 
\end{equation}

\begin{equation}
\label{derivative2}
\displaystyle{\left.\frac{\partial S}{\partial V}\right|_{E,N}}  =
\displaystyle{\frac{P}{T}} 
\end{equation}

\begin{equation}
\label{derivative3}
\displaystyle{\left.\frac{\partial S}{\partial N}\right|_{E,V}}  =
-\displaystyle{\frac{\mu}{T}} .
\end{equation}
We may summarize all these by writing
\[            dS = \frac{1}{T} dE + \frac{P}{T} dV - \frac{\mu}{T} dN  \]
or equivalently
\begin{equation}
\label{differentials}
          dE = T dS - P dV + \mu dN  .
\end{equation}
Starting from the latter equation we see:
\begin{equation}
\label{derivatives1}
\displaystyle{\left.\frac{\partial E}{\partial S}\right|_{V,N}}  =
\displaystyle{T} 
\end{equation}

\begin{equation}
\label{derivatives2}
\displaystyle{\left.\frac{\partial E}{\partial V}\right|_{S,N}} =
\displaystyle{-P}
\end{equation}

\begin{equation}
\label{derivatives3}
\displaystyle{\left.\frac{\partial E}{\partial N}\right|_{S,V}}  =
\displaystyle{\mu} .
\end{equation}

With these definitions, we can start to get a feel for what the
conjugate variables are measuring.  To build intuition, it is useful
to think of the entropy $S$ as roughly the logarithm of the number of
programs whose log runtimes, length and output lie in small ranges $E
\pm \Delta E$, $V \pm \Delta V$ and $N \pm \Delta N$. This is at best
approximately true, but in ordinary thermodynamics this approximation
is commonly employed and yields spectacularly good results.  That is
why in thermodynamics people often say the entropy is the logarithm of
the number of microstates for which the observables $E, V$ and $N$ lie
within a small range of their specified values \cite{Reif}.

If you allow programs to run longer, more of them will halt and give
an answer.  The \textbf{algorithmic temperature}, $T$, is
roughly the number of times you have to double the runtime in order 
to double the number of ways to satisfy the constraints on length and output.

The \textbf{algorithmic pressure}, $P$, measures the tradeoff between 
runtime and length \cite{CSHalting2006}: if you want to keep the number of 
ways to satisfy the constraints constant, then the freedom gained by having 
longer runtimes has to be counterbalanced by shortening the programs.
This is analogous to the pressure of gas in a piston: if you want
to keep the number of microstates of the gas constant, then the freedom 
gained by increasing its energy has to be counterbalanced by decreasing
its volume.  

Finally, the \textbf{algorithmic potential} describes the relation
between log runtime and output: it is a quantitative measure of the
principle that most large outputs must be produced by long programs.

\subsection{Thermodynamic Cycles} \label{cycles}

One of the first applications of thermodynamics was to the analysis
of heat engines.  The underlying mathematics applies equally well
to algorithmic thermodynamics.  Suppose $C$ is a loop in $(T,P,\mu)$ 
space.  Assume we are in a region that can also be coordinatized by the 
variables $E,V,N$.  Then the change in {\bf algorithmic heat} 
around the loop $C$ is defined to be
\[        \Delta Q = \oint_C T dS  .\]
Suppose the loop $C$ bounds a surface $\Sigma$.  Then Stokes' theorem
implies that
\[        \Delta Q = \oint_C T dS  = \int_{\Sigma} dT dS . \]
However, Equation (\ref{differentials}) implies that
\[    dT dS = d(T dS) = d(dE + P dV - \mu dN) = + dP dV - d\mu dN \]
since $d^2 = 0$.  So, we have
\[      \Delta Q = \int_{\Sigma} (dP dV - d\mu dN)  \]
or using Stokes' theorem again
\begin{equation}
\label{loop}
      \Delta Q = \int_C (P dV - \mu dN). 
\end{equation}

In ordinary thermodynamics, $N$ is constant for a heat engine using
gas in a sealed piston.  In this situation we have
\[
      \Delta Q = \int_C P dV  .
\]
This equation says that the change in heat of the gas equals the work
done on the gas---or equivalently, minus the work done \emph{by} the
gas.  So, in algorithmic thermodynamics, let us define $\int_C P dV$
to be the {\bf algorithmic work} done on our ensemble of programs as
we carry it around the loop $C$.  Beware: this concept is unrelated to
`computational work', meaning the amount of computation done by a program 
as it runs.

To see an example of a cycle in algorithmic thermodynamics, consider
the analogue of the heat engine patented by Stoddard in 1919
\cite{Stoddard1919}.  Here we fix $N$ to a constant value and consider 
the following loop in the $PV$ plane:
\begin{center}
  \begin{tikzpicture}
    \draw [->] (-.25,1)--(-.25,4);
    \draw [->] (-.25,1)--(4,1);
    \draw [->] (1,2)--(1,3.5);
    \draw [->] (1,3.5) .. controls (1.5, 2.5) and (1.75, 2.5) .. (3,2);
    \draw [->] (3,2)--(3,1.5);
    \draw [->] (3,1.5).. controls (1.5,1.75) and (1.75, 1.75) .. (1,2);
    \node at (-.25,2) [left] {$P$};
    \node at (2,1) [below] {$V$};
    \node at (1,2.75) [left] {1};
    \node at (2,2.5) [above] {2};
    \node at (3,1.75) [right] {3};
    \node at (2,1.7) [below] {4};
    \begin{scope}[font=\fontsize{5}{5}\selectfont]
      \node at (1,2) [below left] {$(P_1, V_1)$};
      \node at (1,3.5) [above left] {$(P_2, V_1)$};
      \node at (3,2) [above right] {$(P_3, V_2)$};
      \node at (3,1.5) [below right] {$(P_4, V_2)$};
    \end{scope}
  \end{tikzpicture}
\end{center}
We start with an ensemble with algorithmic pressure $P_1$ and mean
length $V_1$.  We then trace out a loop built from four parts:

\begin{enumerate}
  \item 
  \emph{Isometric}. We increase the pressure from $P_1$ to $P_2$ while keeping
  the mean length constant.  No algorithmic work is done on the ensemble
  of programs during this step.
  \item
  \emph{Isentropic.} We increase the length from $V_1$ to $V_2$ while keeping
  the number of halting programs constant.  High pressure means 
  that we're operating in a range of runtimes where if we increase
  the length a little bit, many more programs halt.  In order
  to keep the number of halting programs constant,
  we need to shorten the runtime significantly.  As we gradually
  increase the length and lower the runtime, the pressure drops to $P_3$.
  The total difference in log runtime is the algorithmic 
  work done on the ensemble during this step.
  \item
  \emph{Isometric.} Now we decrease the pressure from $P_3$ to $P_4$ while
  keeping the length constant.  No algorithmic work is done during this step.
  \item
  \emph{Isentropic.} Finally, we decrease the length from $V_2$ back to $V_1$
  while keeping the number of halting programs constant.
  Since we're at low pressure, we need only increase the
  runtime a little.  As we gradually decrease the length
  and increase the runtime, the pressure rises slightly
  back to $P_1$.  The total increase in log runtime is minus the algorithmic
  work done on the ensemble of programs during this step.
\end{enumerate}
The total algorithmic work done on the ensemble per cycle is the 
difference in log runtimes between steps 2 and 4.

\subsection{Further Relations} \label{maxwell}

From the elementary thermodynamic relations in Section 
\ref{elementary}, we can derive various others.  
For example, the so-called `Maxwell relations' are obtained
by computing the second derivatives of thermodynamic quantities in two
different orders and then applying the basic derivative relations,
Equations (\ref{derivatives1}-\ref{derivatives3}).  While 
trivial to prove, these relations say some things about algorithmic 
thermodynamics which may not seem intuitively obvious.

We give just one example here.  Since mixed partials commute, we have:
\[  \left.\frac{\partial^2 E}{\partial V \partial S}\right|_N = 
\left.\frac{\partial^2 E}{\partial S \partial V}\right|_N . \]
Using Equation (\ref{derivatives1}), the left side can be computed
as follows:
\[  \left.\frac{\partial^2 E}{\partial V \partial S}\right|_N = 
\left.\frac{\partial}{\partial V}\right|_{S,N}
\left.\frac{\partial E}{\partial S}\right|_{V,N} = 
\left.\frac{\partial T}{\partial V}\right|_{S,N}  \]
Similarly, we can compute the right side
with the help of Equation (\ref{derivatives2}):
\[  \left.\frac{\partial^2 E}{\partial S \partial V}\right|_N = 
\left.\frac{\partial}{\partial S}\right|_{V,N}
\left.\frac{\partial E}{\partial V}\right|_{S,N} = 
-\left.\frac{\partial P}{\partial S}\right|_{V,N} . \]
As a result, we obtain:
\[  \left.\frac{\partial T}{\partial V}\right|_{S,N}  =
-\left.\frac{\partial P}{\partial S}\right|_{V,N} . \]

We can also derive interesting relations involving derivatives of
the partition function.  These become more manageable if 
we rewrite the partition function in 
terms of the conjugate variables of the observables $E, V$, and $N$:
\begin{equation}
\label{new.variables}
   \beta = \frac{1}{T} , \quad
 \gamma =  \frac{P}{T},  \quad
\delta = -\frac{\mu}{T}.
\end{equation}
Then we have
\[     Z = \sum_{x \in X} e^{-\beta E(x) -\gamma V(x) - \delta N(x)} \]

Simple calculations, standard in statistical mechanics 
\cite{Reif}, then allow us to compute the mean values of 
observables as derivatives of the logarithm of $Z$ 
with respect to their conjugate variables.    Here
let us revert to using overlines to denote mean values:
\[
\begin{array}{ccc}
   \overline{E} &= 
   \displaystyle{\sum_{x \in X} p(x) \, E(x)} &=
- \displaystyle{\frac{\partial}{\partial \beta} \ln Z}  \\  
\\
   \overline{V} &=
   \displaystyle{\sum_{x \in X} p(x) \, V(x)} &=
- \displaystyle{\frac{\partial}{\partial \gamma} \ln Z}  \\  
\\
   \overline{N} &=
   \displaystyle{\sum_{x \in X} p(x) \, N(x)} &=
- \displaystyle{\frac{\partial}{\partial \delta} \ln Z}  
\end{array}
\]
We can go further and compute the variance of these observables
using second derivatives:
\[       
\begin{array}{ccc}
   {(\Delta E)^2} &=
\displaystyle{\sum_{x \in X} p(x) (E(x)^2 - \overline{E}^2)} &=
\displaystyle{ \frac{\partial^2}{\partial^2 \beta} \ln Z } 
\end{array}
\]
and similarly for $V$ and $N$.  Higher moments of $E, V$ and 
$N$ can be computed by taking higher derivatives
of $\ln Z$.  

\subsection{Convergence}

So far we have postponed the crucial question of convergence:
for which values of $T,P$ and $\mu$ does the partition function 
$Z$ converge?  For this it is most convenient to treat $Z$ as
a function of the variables $\beta, \gamma$ and $\delta$
introduced in Equation (\ref{new.variables}).  For which 
values of $\beta, \gamma$ and $\delta$ does the partition
function converge?   

First, when $\beta = \gamma = \delta = 0,$ the contribution of each program 
is 1.   Since there are infinitely many halting programs, $Z(0,0,0)$ does 
not converge.

Second, when $\beta = 0, \gamma = \ln 2,$ and $\delta = 0,$ the partition
function converges to Chaitin's number
\[ \Omega = \sum_{x \in X} 2^{-V(x)}.  \]
To see that the partition function converges in this case, consider 
this mapping of strings to segments of the unit interval: 
\begin{center}
  \begin{tikzpicture}
    \draw (0,2)--(0,0);
    \draw (8,2)--(8,0);
    \draw (4,1.5)--(4,0);
    \draw (2,1)--(2,0);
    \draw (6,1)--(6,0);
    \draw (1,.5)--(1,0);
    \draw (3,.5)--(3,0);
    \draw (5,.5)--(5,0);
    \draw (7,.5)--(7,0);
    \node at (4,1.75) {empty};
    \node at (2,1.25) {0};
    \node at (6,1.25) {1};
    \node at (1,0.75) {00};
    \node at (3,0.75) {01};
    \node at (5,0.75) {10};
    \node at (7,0.75) {11};
    \node at (0.5,.25) {000};
    \node at (1.5,.25) {001};
    \node at (2.5,.25) {010};
    \node at (3.5,.25) {011};
    \node at (4.5,.25) {100};
    \node at (5.5,.25) {101};
    \node at (6.5,.25) {110};
    \node at (7.5,.25) {111};
    \node at (4,-0.25) {$\vdots$};
    \foreach \y in {0,1,2,3,4} \draw (0,\y/2)--(8,\y/2);
  \end{tikzpicture}
\end{center}

Each segment consists of all the real numbers whose binary
expansion begins with that string; for example, the set of
real numbers whose binary expansion begins $0.101\ldots$
is [0.101, 0.110) and has measure $2^{-|101|} = 2^{-3} = 1/8.$
Since the set of halting programs for our universal machine
is prefix-free, we never count any segment more than once,
so the sum of all the segments corresponding to halting 
programs is at most 1.

Third, Tadaki has shown \cite{Tadaki2002} that the expression  
\[ \sum_{x \in X} e^{-\gamma V(x)} \]
converges for $\gamma \ge \ln 2$ but diverges for $\gamma < \ln 2.$ 
It follows that $Z(\beta, \gamma, \delta)$
converges whenever $\gamma \ge \ln 2$ and $\beta, \delta \ge 0$.  

Fourth, when $\beta>0$ and $\gamma=\delta=0,$ convergence depends 
on the machine.  There are machines where infinitely many programs 
halt immediately.  For these, $Z(\beta,0,0)$ does not converge.
However, there are also machines where program $x$ takes at least 
$V(x)$ steps to halt; for these machines $Z(\beta,0,0)$ will converge 
when $\beta \ge \ln 2.$  Other machines take much longer to run.
For these, $Z(\beta,0,0)$ will converge for even smaller values of $\beta$.  

Fifth and finally, when $\beta = \gamma = 0$ and $\delta > 0$, 
$Z(\beta,\gamma,\delta)$ fails to converge, since there are infinitely many 
programs that halt and output 0.

\subsection{Computability} \label{computability}

Even when the partition function $Z$ converges, it may not be 
computable.  The theory of computable real numbers was independently
introduced by Church, Post, and Turing, and later blossomed into 
the field of computable analysis \cite{PourElRichards}.
We will only need the basic definition: a real number $a$ is {\bf 
computable} if there is a recursive function that maps any
natural number $n > 0$ to an integer $f(n)$ such that
\[              \frac{f(n)}{n} \le a \le \frac{f(n)+1}{n}  .\]
In other words, for any $n > 0$, we can compute a rational number
that approximates $a$ with an error of at most $1/n$.  This definition
can be formulated in various other equivalent ways: for example,
the computability of binary digits.

Chaitin \cite{Chaitin1975} proved that the number
\[     \Omega = Z(0,\ln 2, 0)  \]
is uncomputable.  In fact, he showed that for any universal 
machine, the values of all but finitely many bits of $\Omega$ are not 
only uncomputable, but random: knowing the value of some of them 
tells you nothing about the rest.  They're independent, like separate 
flips of a fair coin.

More generally, for any computable number $\gamma \ge \ln 2$, 
$Z(0,\gamma,0)$ is `partially random' in the sense of 
Tadaki \cite{CST2004,Tadaki2002}.  
This deserves a word of explanation.  A fixed formal system with 
finitely many axioms can only prove finitely many bits of 
$Z(0,\gamma,0)$ have the values they do; after that, 
one has to add more axioms or rules to the system to make any progress.  
The number $\Omega$ is completely random in the following sense:
for each bit of axiom or rule one adds, one can prove at most one 
more bit of its binary expansion has the value it does.  So, the most 
efficient way to prove the values of these bits is simply to add 
them as axioms!  But for $Z(0,\gamma,0)$ with $\gamma > \ln 2$,
the ratio of bits of axiom per bits of sequence is less than than 1.
In fact, Tadaki showed that for any computable $\gamma \ge \ln 2$, 
the ratio can be reduced to exactly $(\ln 2)/\gamma$.

On the other hand, $Z(\beta,\gamma,\delta)$ is computable for 
all computable real numbers $\beta > 0$, $\gamma \ge \ln 2$ and 
$\delta \ge 0$.   The reason is that $\beta > 0$ exponentially 
suppresses the contribution of machines with long runtimes, 
eliminating the problem posed by the undecidability of the
halting problem.  The fundamental insight here is due to 
Levin \cite{Levin1973}.  His idea was to `dovetail' all 
programs: on turn $n$, run each of the first $n$ programs a 
single step and look to see which ones have halted.  As they 
halt, add their contribution to the running estimate of $Z$.  
For any $k\ge 0$ and turn $t\ge 0$, let $k_t$ be the location 
of the first zero bit after position $k$ in the estimation of $Z$.  
Then because $-\beta E(x)$ is a monotonically decreasing function 
of the runtime and decreases faster than $k_t$, there will be a 
time step where the total contribution of all the programs that 
have not halted yet is less than $2^{-k_t}$.

\section{Conclusions}

There are many further directions to explore.  Here we mention just
three.  First, as already mentioned, the `Kolmogorov complexity' 
\cite{Kolmogorov1965} of a number $n$ is the number of bits in the 
shortest program that produces $n$ as output.   
However, a very short program that runs for a million years before
giving an answer is not very practical.  To address this problem,
the \textbf{Levin complexity} \cite{Levin1974} of $n$ is defined
using the program's length plus the logarithm of its runtime, again 
minimized over all programs that produce $n$ as output.  Unlike 
the Kolmogorov complexity, the Levin complexity is computable.  But 
like the Kolmogorov complexity, the Levin complexity can be seen 
as a \emph{relative entropy}---at least, up to some error bounded 
by a constant.   The only difference is that now we compute this 
entropy relative to a different probability measure: instead of using 
the Gibbs distribution at infinite algorithmic temperature, we drop 
the temperature to $\ln 2$.  Indeed, the Kolmogorov and Levin 
complexities are just two examples from a continuum of options.
By adjusting the algorithmic pressure and temperature, we get 
complexities involving other linear combinations of length and log 
runtime.  The same formalism works for complexities involving 
other observables: for example, the maximum amount of memory 
the program uses while running.

Second, instead of considering Turing machines that output
a single natural number, we can consider machines that output a finite
list of natural numbers $(N_1, \ldots, N_j);$ we can treat these as
populations of different ``chemical species'' and define
algorithmic potentials for each of them.  Processes analogous to 
chemical reactions are paths through this space that preserve 
certain invariants of the lists.  With chemical reactions we can
consider things like internal combustion cycles.

Finally, in ordinary thermodynamics the partition function $Z$ is
simply a number after we fix values of the conjugate variables.  The
same is true in algorithmic thermodynamics.  However, in algorithmic
thermodynamics, it is natural to express this number in binary
and inquire about the algorithmic entropy of the first $n$ bits.  For
example, we have seen that for suitable values of temperature, 
pressure and chemical potential, $Z$ is Chaitin's number $\Omega$.  
For each universal machine there exists a constant $c$ such 
that the first $n$ bits of the number $\Omega$ have at least 
$n - c$ bits of algorithmic entropy with respect to that 
machine.  Tadaki \cite{Tadaki2002} generalized this computation to 
other cases.

So, \textit{in algorithmic thermodynamics, the partition
function itself has nontrivial entropy}.  Tadaki has shown that
the same is true for algorithmic pressure (which in his analogy
he calls `temperature').   This reflects the self-referential 
nature of computation.  It would be worthwhile to understand this 
more deeply.

\subsection*{Acknowledgements} 
We thank Leonid Levin and the denizens of the $n$-Category Caf\'e for
useful comments.  MS thanks Cristian Calude for many discussions 
of algorithmic information theory.  JB thanks Bruce Smith for 
discussions on relative entropy.  He also thanks Mark Smith for 
conversations on physics and information theory, as well as for 
giving him a copy of Reif's {\sl Fundamentals of Statistical and 
Thermal Physics}.

\end{document}